# A Microscopic Estimation of the Splay Elastic Constant $K_{11}$ for Nematic Liquid Crystals

Tien-Lun Ting and Tsung-Hsien Lin[1]

[1]Department of Photonics, National Sun Yat-sen University, No. 70 Lien-hai Rd. Kaohsiung, Taiwan 804201, R.O.C.

A theory of the relationship between splay elastic constant $K_{11}$ and permanent molecular moment of liquid crystals is developed. The theory gives a microscopic estimation of $K_{11}$ with a correct order of magnitude by calculating the electrostatic energy density between the adjacent molecules.

## I. INTRODUCTION

Since Frank proposed the general theory of curvature-elasticity in the molecularly uniaxial liquid crystals[1], research of elastic constants of liquid crystals have been reported. Among these studies, most of them were about measuring the elastic constants by experiment. Using these measured elastic constants one can calculate the Frank free energy of a nematic liquid crystal with deformation. It is not exaggerative to say that the entire LCD industry depends greatly upon these studies to simulate the possible alignment of LC molecules under different applied electric fields so as to deliver better performances. Other than the measurement, several studies gave excellent statistical model to calculate the elastic constants[2-6]. There is, however, a question left to be answered: What is the origin of the elasticity? Among the fundamental forces of nature, it is reasonable to conclude that the elasticity originates from the electromagnetic force. If so, what is the mechanism? In this paper, a microscopic model is developed. The nematic liquid crystals are treated as permanent molecular dipole moments. The interaction of these dipoles will create electrostatic energy between them. As deformation occurs, such electrostatic energy will be altered and hence the molecules perceive a force, which is regarded as elasticity. By calculating the change of electrostatic energy, one can estimate $K_{11}$. Meanwhile, the orientation polarization caused by permanent dipole is calculated and compared to the electric polarization caused by induced dipole. The results show that it is possible to attribute the elasticity to the permanent dipole since the induced one is much weaker. Besides, the electric polarization still dominates when applying electric field to the nematic liquid crystals. These calculations agree well with the phenomena in the real world.

## II. MICROSCOPIC ESTIMATION OF $K_{11}$

The idea of microscopic estimation of splay elastic constant is very simple. Each nematic liquid crystal molecule is considered as a permanent dipole. The electric field from these dipoles has electrostatic energy. When deformation occurs, the electrostatic energy stored between liquid crystal molecules change and hence a force is perceived by molecules. This force is the origin of elasticity. If the change of the electrostatic energy is carefully calculated, one can derive the elastic constants of a nematic liquid crystal. Fig. 1 is the schematic diagram of a dipole moment $p_0$, which equals $q \times d$. The electric field at point $A$, whose distance from the origin $r$ is much larger than $d$, can be written as (1) and (2) corresponding to $z$ and $xy$-plane component, respectively.

$$E_z|_A = \frac{p_0}{4\pi\varepsilon_0}\left(\frac{3\cos^2\theta - 1}{r^3}\right) \tag{1}$$

$$E_{xy}|_A = \frac{p_0}{4\pi\varepsilon_0}\left(\frac{3\cos\theta\sin\theta}{r^3}\right) \tag{2}$$

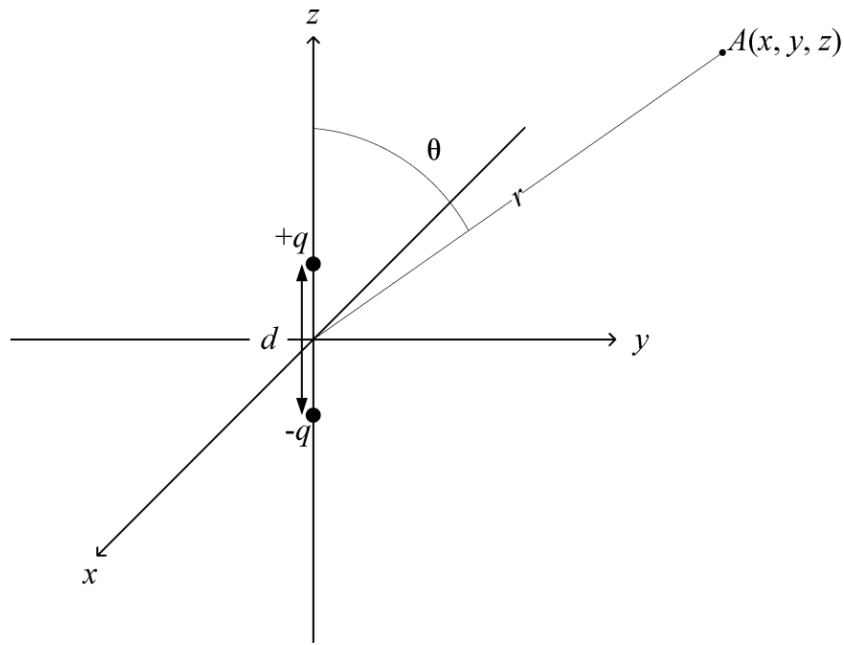

Fig. 1. Schematic diagram of a dipole moment

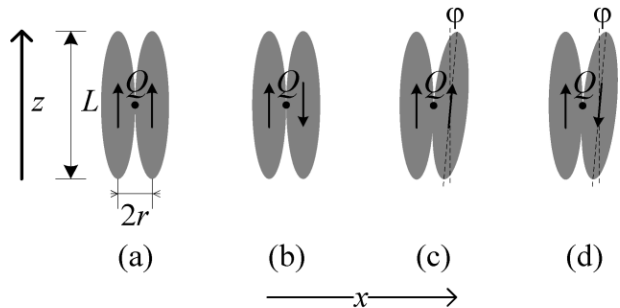

Fig. 2. Schematic diagram of liquid crystal molecules with permanent dipoles: (a) parallel alignment, (b) antiparallel alignment, (c) parallel alignment with a small splay deformation, and (d) antiparallel alignment with a small splay deformation

As liquid crystal molecules are treated as permanent dipoles, one can simplify the analysis by merely considering the interaction between two neighboring molecules. Fig. 2(a) and (b) are the two possible ways for two adjacent dipoles to align without deformation. For Fig. 2(a) the electric field at point $Q$ is the superposition of two parallel dipoles as shown in (3). The superscript $p$ means parallel alignment. Only $E_z$ exists since $E_{xy}$ is completely cancelled by each other. This means that the electrostatic energy density at point $Q$ of Fig. 2(a) can be expressed by (4). As for the antiparallel alignment of Fig. 2(b), no net electric field arises at $Q$, and thus the electrostatic energy density is zero.

$$E_z^p\Big|_Q = \frac{p_0}{4\pi\varepsilon_0}\left[\frac{3\cos^2\left(\frac{\pi}{2}\right)-1}{r^3}\right] + \frac{p_0}{4\pi\varepsilon_0}\left[\frac{3\cos^2\left(-\frac{\pi}{2}\right)-1}{r^3}\right] = -\frac{p_0}{2\pi\varepsilon_0 r^3} \tag{3}$$

$$u_0^p\Big|_Q = \frac{\varepsilon_0}{2}\boldsymbol{E}\cdot\boldsymbol{E} = \frac{p_0^2}{8\varepsilon_0\pi^2 r^6} \tag{4}$$

When a small splay deformation occurs, the electric fields at $Q$ will definitely be different from their own original states. Fig. 2(c) and (d) represent the parallel and the antiparallel alignment with splay deformation. For Fig. 2(c), the $z$-component e-field is characterized by (5) and $xy$-plane by (6). When the deformation angle φ is small enough, the electrostatic energy density becomes (7), and its difference from (4) is shown in (8). The sign of (8) being negative indicates that when two dipoles align in parallel, they tend to splay to lower the energy of the system. By the same method one can derive the energy density of the antiparallel with splay, and it will increase by the same amount.

$$E_z^p(\varphi)\Big|_Q = \frac{p_0}{4\pi\varepsilon_0}\left[\frac{3\cos^2\left(\frac{\pi}{2}\right)-1}{r^3} + \frac{3\cos^2\left(\frac{3\pi}{2}-\varphi\right)-1}{r^3}\cos\varphi - \frac{3\cos\left(\frac{3\pi}{2}-\varphi\right)\sin\left(\frac{3\pi}{2}-\varphi\right)}{r^3}\sin\varphi\right] \tag{5}$$

$$E_{xy}^p(\varphi)\Big|_Q = \frac{p_0}{4\pi\varepsilon_0}\left[\frac{3\cos^2\left(\frac{\pi}{2}\right)-1}{r^3}\sin\varphi + \frac{3\cos\left(\frac{3\pi}{2}-\varphi\right)\sin\left(\frac{3\pi}{2}-\varphi\right)}{r^3}\cos\varphi\right] \tag{6}$$

$$u_1^p\Big|_Q = \frac{p_0^2}{8\varepsilon_0\pi^2 r^6}\frac{2+2\cos\varphi+3\sin^2\varphi}{4} \tag{7}$$

$$\begin{aligned}\Delta u^p\Big|_Q &= \frac{p_0^2}{8\varepsilon_0\pi^2 r^6}\left(2-2\cos\varphi-3\sin^2\varphi\right) = \frac{p^2}{8\varepsilon_0\pi^2 r^6}\left(2-2\left(1-\varphi^2\right)-3\varphi^2\right)+O\left(\varphi^3\right)\\ &\approx -\frac{p^2\varphi^2}{8\varepsilon_0\pi^2 r^6} = -\Delta u^a\Big|_Q\end{aligned} \tag{8}$$

This energy density change corresponds to the elastic energy density of the splay deformation. Without loss of generality, one can choose the *x*-direction along the direction with deformation, and thus the elastic energy density at *Q* can be expressed as (9), where $n_x$ indicates the *x*-component of the unit vector representing the direction of the dipole. Since the electrostatic energy density is not identical everywhere between molecules, to estimate $K_{11}$ the length of molecules, *L*, must be considered. By calculating the average electrostatic energy density, $K_{11}$ can be written as (10). If the permanent molecular moment is around 0.1 electron×Å and the molecular dimensions are a 5-Å diameter by a 20-Å length, this gives a $K_{11}$ of 14.43 pN with a correct order of magnitude.

$$u_{elastic} = \frac{K_{11}}{2}\left(\frac{\partial n_x}{\partial x}\right)^2 = \frac{K_{11}}{2}\left(\frac{\sin\varphi}{\Delta x}\right)^2 \approx \frac{K_{11}\varphi^2}{8r^2} \tag{9}$$

$$K_{11} = \frac{4r^2\varepsilon_0 \int_{-\frac{L}{2}}^{\frac{L}{2}}(E_1^2-E_0^2)dz}{L\varphi^2} \tag{10}$$

### III. POLARIZATION FROM PERMANENT AND INDUCED DIPOLE

The previous section states that the splay elastic constant comes from the interaction of the permanent molecular dipole moments. This will naturally raise a question: Does the induced molecular dipole moment caused by the applied electric field contribute to $K_{11}$ as well? To answer this question, we need to know how large the induced dipole is. The moment induced by electric field **E** can be expressed by (11), where $\chi$ is the electric susceptibility and *N* is the number density of the liquid crystal. If the liquid crystal has a molecular weight and density of 700 and 0.9 g/cm$^3$, respectively, it leads to an *N* of $7.74\times10^{26}$ m$^{-3}$. When *E* of $10^6$ V/m is applied and $\chi$ is assumed to be 6, the induced dipole will be 0.0043 electron×Å, which is much smaller than the permanent dipole in the previous section. Meanwhile, $K_{11}$ is proportional to the square of the dipole moment. This means that the contribution to $K_{11}$ from the permanent dipole is 500 times larger than that from the induced dipole. Therefore, one can be certain that the splay elastic constant is hardly influenced by the applied electric field in general.

$$\boldsymbol{p}_i = \frac{\varepsilon_0 \chi E}{N} \tag{11}$$

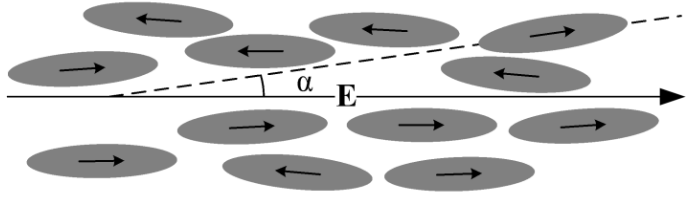

Fig. 3. The schematic diagram of nematic liquid crystal in an electric field

There is another question to be answered if one would like to consider a liquid crystal molecule as a permanent dipole. What is the polarization field caused by such molecular moment? When an electric field is applied, the permanent molecular moment will be rearranged, and thus a net polarization field may be formed. Such polarization is known as the orientation polarization. If the nematic liquid crystal molecules are confined within the solid angle $2\pi(1-\cos\alpha)$, where α is the colatitude from the electric field as shown in Fig. 3, the number density $N$ in the previous paragraph shall follow (12). Since the potential energy of a dipole in electric field equals $-pE\cos\theta$, the number of molecules at θ per unit solid angle, $n(\theta)$, can be expressed by (13). By substituting $n(\theta)$ in (12) by (13), one can have $n(\theta)$ expressed by $N$. Eventually the orientation polarization $P_o$ can be summed over the whole angle distribution by (14), and its direction will be along the electric field because more molecules orient along than against the field. Again, we can put real numbers into the equations to check how large this orientation polarization is. When α is 90°, $P_o$ is $1.6 \times 10^{-7}$ C/m², and it becomes $4.8 \times 10^{-7}$ C/m² while α is 5°. Either of them is much smaller than the electric polarization $\varepsilon_0 \chi E$, which is $5.3 \times 10^{-5}$ C/m² with the same numbers used above. The dominant polarization in a nematic liquid crystal is still the electric polarization.

$$N = \int_0^{2\pi} \int_0^{\alpha} n(\theta) \sin\theta \, d\theta d\phi + \int_0^{2\pi} \int_{\pi-\alpha}^{\pi} n(\theta) \sin\theta \, d\theta d\phi \tag{12}$$

$$n(\theta) = n_0 \exp\left(\frac{-U}{kT}\right) = n_0 \exp\left(\frac{-pE\cos\theta}{kT}\right) \approx n_0 \left(1 + \frac{pE\cos\theta}{kT}\right) \tag{13}$$

$$P_o = p\left(\int_0^{\alpha} n(\theta) \cos\theta \, 2\pi \sin\theta \, d\theta + \int_{\pi-\alpha}^{\pi} n(\theta) \cos\theta \, 2\pi \sin\theta \, d\theta\right) \tag{14}$$

**IV. CONCLUSION**

A microscopic estimation of $K_{11}$ is demonstrated. The estimation gives the correct order of magnitude of $K_{11}$ for nematic liquid crystals and shows that the $K_{11}$ will be influenced by the permanent dipole moment, the aspect ratio of liquid crystal molecules, and the distance between molecules. It is shown that the elastic constant will not be altered by the applied electric field because the induced dipole moment is much smaller than the permanent one. It is also shown that the orientation polarization caused by the permanent dipole moment is much weaker than the polarization caused by the applied electric field. Such interpretation can lead to the conclusion that the microscopic permanent dipole moment is the origin of the elasticity between the nematic liquid crystal molecules. As for the non-polarized liquid crystal molecules, one can consider it as a quadrupole instead of a permanent dipole. The electric field generated by quadrupole decreases with distance to the fourth power and is much weaker than dipole-generated one. This implies the elastic constants of a non-polarized liquid crystal molecule will be smaller than that of a polarized one.

**REFERENCES**

[1] F. C. Frank, Discuss. Faraday Soc. **25**, 19 (1958).

[2] T. C. Lubensky, Phys. Lett. A **33**, 202 (1970).

[3] R. G. Priest, Phys. Rev. A **7**, 720 (1973).

[4] J. P. Straley, Phys. Rev. A **8**, 2181 (1973).

[5] A. Poniewiersky and J. Stecki, Phys. Rev. A **25**, 2368 (1982).

[6] S.-D. Lee and R. B. Meyer, J. Chem. Phys **84**, 3443 (1986).